\documentclass[journal]{IEEEtran}

\IEEEoverridecommandlockouts

\usepackage{ieeefig}
\usepackage{latexsym}
\usepackage{graphicx}
\usepackage{epsfig}
\usepackage{subfigure}
\usepackage{array}
\usepackage{amsmath}
\usepackage{amssymb}
\usepackage{color}
\usepackage{amsthm}
\usepackage{enumerate}
\theoremstyle{plain} 

\theoremstyle{definition}

\newcommand{\bxi} {\boldsymbol{\xi}}

\newcommand{\bx}{{\bf x}}
\newcommand{\by}{{\bf y}}

\def\bal#1\eal{\begin{align}#1\end{align}}
\newcommand{\bp} {\begin{proof}}
\newcommand{\ep} {\end{proof}}

\newcommand{{\Rb}} {\right)}

\newcommand{{\Rf}} {\right\}}

\begin{document}

\title{Simple Closed-Form Approximations for Achievable Information Rates of Coded Modulation Systems}
\author{Maria Urlea and Sergey Loyka

\vspace*{-1.5\baselineskip}

\thanks{The authors are with the School of Electrical Engineering and Computer Science, University of Ottawa, Ontario, Canada, K1N 6N5, e-mail: sergey.loyka@uottawa.ca}
}

\maketitle

\begin{abstract}
The intuitive sphere-packing argument is used to obtain analytically-tractable closed-form approximations for achievable information rates of coded modulation transmission systems, for which only analytically-intractable expressions are available in the literature. These approximations provide a number of insights, possess useful properties and facilitate design/optimization of such systems. They apply to constellations of various cardinalities (including large ones), are simple yet reasonably accurate over the whole signal-to-noise ratio range, and compare favorably to the achieved rates of recent state-of-the art experiments.
\end{abstract}


%
\section{Introduction}
\label{sec:introduction}

The growing demand for higher transmission rates and higher spectral efficiency along with the development of powerful capacity-approaching codes has recently stimulated significant interest in coded-modulation optical transmission systems, with multi-level modulation and capacity-approaching codes \cite{Karlsson-17}\cite{Cho-19}. For such systems, an adequate performance metric is achievable information rate (AIR), which can be expressed as mutual information (MI) or generalized mutual information (GMI), depending on the specifics of decoding/demodulation algorithms \cite{Fehenberger-15}\cite{Alvarado-18}. They have been widely used for analysis, optimization and design of various communication systems, including optical fiber systems.

While achievable information rates have been studied in the information-theoretic literature for multi-level modulation formats over the additive white Gaussian noise (AWGN) channel \cite{ungerboeck}-\cite{ozarow}, the corresponding expressions cannot be evaluated analytically in a closed form, even for the simplest AWGN channel, since they include a number of analytically-intractable integrals over infinite intervals. Numerical integration (e.g., via Monte-Carlo approach) remains the only alternative \cite{Alvarado-18}\cite{ungerboeck}. This makes it difficult to obtain insights  and also to perform system design and optimization (for example, to find an optimal power allocation in a multi-stream transmission system, as in e.g. \cite{lozano}).

In this paper, we present analytically-tractable closed-form approximations to achievable information rates for $M$-PAM and $M$-QAM constellations (modulation) on the AWGN channel (possibly including nonlinear interference in optical fiber modeled as additional Gaussian noise \cite{Poggiolini-14}) using the intuitive sphere-packing argument. These approximations are simple yet reasonably accurate over the whole signal-to-noise ratio (SNR) range (not just at high or low SNR) and for various constellation cardinalities $M$ (including large ones). Comparison to the back-to-back (B2B) or long-haul  end-to-end (E2E) rates of recent state-of-the-art systems/experiments shows that the proposed approximations are in fact more accurate than the ideal AIR (via MI or GMI) since the latter ignores many imperfections and limitations of real-world systems (e.g. non-zero overhead of realistic codes, guard bands, etc.).

The proposed approximations also provide a number of insights unavailable from the exact (MI-based) expressions. In particular, they allow one to obtain the minimum $M$ required to approach closely the (modulation-unconstrained) AWGN channel capacity without using unnecessarily large constellations. The derivation of these approximations offers an additional insight into a mechanism causing a rate loss in the modulation-constrained system: while the rate loss is additive at low SNR, it is multiplicative, i.e., much more pronounced, at high SNR. The obtained approximations possess useful properties: they are differentiable and concave in the SNR so that the respective optimization problems are convex and thus all powerful tools of convex optimization (see e.g. \cite{boyd}) can be used to solve them. The main contributions of this paper are the approximations of the modulation-constrained achievable information rates of $M$-PAM and $M$-QAM constellations in \eqref{eq.CM.approx}, \eqref{eq:MQAM}, and \eqref{eq.C2a}, from which the minimum number of constellation points to approach closely the channel capacity (without significant rate loss due to  limited $M$) can be found as in \eqref{eq.Mmin} and \eqref{eq.Mmin.QAM}.

\section{Channel Model and Information Rates}
\label{sec:syst}

\begin{figure}[t]
    \centerline{\includegraphics[width=3.6in]{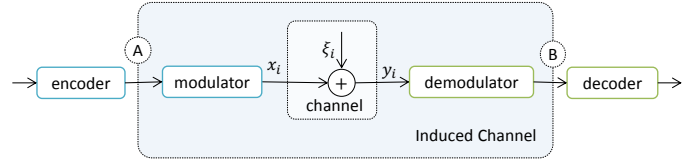}}
    \caption{System Model and Induced Channel.}
    \label{fig:1}
\end{figure}

Following the widely-accepted approach, we use the Gaussian channel model, which  has been extensively used in many areas of digital communications \cite{barry}. This model was also demonstrated (via simulations and also experimentally) to be sufficiently accurate for the optical fiber channel in many cases of practical interest (uncompensated transmission in the low-to-moderate nonlinearity regime) and offers a good balance of accuracy and simplicity \cite{Poggiolini-14}\cite{Eriksson-16}. The system block diagram is shown in Fig. \ref{fig:1}, which includes an encoder, a modulator, a channel, a demodulator and a decoder \cite{barry}. The function of the encoder is to protect the transmitted message against channel-induced errors using forward error correction and the function of the modulator is to transform the encoded signal to a form suitable for transmission over the communication channel. The baseband discrete-time channel model (after matched filtering  and sampling) is
\begin{equation}
	y_i =  x_i + \xi_i,
	\label{eq:comSys}
\end{equation}
where $x_i$ and $y_i$ are transmitted and received (real-valued) symbols respectively at time $i$, $\xi_i$ is the additive white Gaussian noise. The capacity of this channel, i.e. the maximum achievable rate under the reliability criterion (arbitrary small error probability) and the power constraint, is \cite{blahut}
\bal
\label{eq.C}
C = \frac{1}{2} \log (1+\gamma) \ \mbox{[bit/symbol]},
\eal
where $\gamma = \sigma_x^2/\sigma_0^2$ is the SNR,  $\sigma_x^2$ and $\sigma_0^2$ are the signal and noise power and all logarithms are base-2; $\sigma_0^2$ can also include nonlinear interference in optical fiber when modeled as Gaussian noise, see e.g. \cite{Poggiolini-14}.

In practical systems, digital modulator makes use of a certain number of signals conveniently represented via a constellation (in the signal space) \cite{barry}. Due to implementation complexity, the constellation cardinality (i.e. the number of points/signals) is limited to $M$. On the other hand, recent development and adoption of powerful capacity-approaching codes allows one to build nearly-optimal encoders and decoders. Consequently, modulation-constrained achievable information rates are widely used as a performance metric \cite{Karlsson-17}-\cite{Alvarado-18}. From the information-theoretic perspective, the part of the system between the encoder and the decoder can be interpreted as an \textit{induced channel}, which includes the actual channel and the constrained (fixed) modulator/demodulator. This is illustrated in Fig. \ref{fig:1}, where the induced channel is between A and B. Assuming symbol-wise decoding, the achievable information rate of this system is the mutual information of the induced channel. This quantity has been studied in the information-theoretic literature \cite{ungerboeck}-\cite{ozarow}.  For the real-valued $M$-PAM constellation under equiprobable signalling, the per-symbol MI $C_M$ is \cite{blahut}
\bal
\label{eq.CM}
	C_M &= \log M \\ \notag
&-\sum_j{\frac{1}{M\sqrt{2\pi\sigma_0^2}} \displaystyle\int_{-\infty}^{\infty}{e^{-\frac{z^2}{2\sigma_0^2}}
\log{\sum_i{e^{-\frac{d_{ij}^2}{2\sigma_0^2}}e^{-\frac{zd_{ij}}{\sigma_0^2}}dz}}}},
\eal
where $d_{ij}=a_i-a_j$ so that $|d_{ij}|$ is the distance between constellation points $a_i$ and $a_j$; the summation is over all points. Unfortunately, the integrals over infinite intervals cannot be evaluated in closed-form, which makes it difficult to obtain insights and to use it in design/optimization process. Numerical evaluation of these integrals can be also difficult, especially when constellation cardinality is large and/or real time evaluation is required (e.g. to compute optimal power allocation for multi-stream transmission system, see e.g. \cite{lozano}).

From the information-theoretic perspective, achieving the channel capacity requires selecting the best encoder/decoder and modulator/demodulator while a modulation-constrained achievable information rate implies the optimization of the encoder/decoder only while the modulator is fixed, so that $C_M \le C$. On the other hand, the AIR cannot exceed the entropy of the corresponding constellation \cite{blahut}, $C_M \le \log M$, so that $C_M$ can be upper-bounded as follows
\bal
\label{eq.CM.bound}
C_M \le \min(C, \log M).
\eal
While the former two bounds can be rather loose individually (at high and low SNR respectively), the latter is significantly tighter when considered over the whole SNR range (see Fig. 4). To obtain further insights and to overcome the above-mentioned difficulties, we present below a simple closed-form approximation for $C_M$.

\section{AIR via the Sphere Packing Argument}
\label{sec:Sphere Packing Argument}

While a rigorous information-theoretic derivation of \eqref{eq.C} is available and well-known \cite{blahut}, the sphere-packing argument, which originates back to Shannon, provides a more intuitive understanding as to why this expression holds \cite{wosencraft}. Since our approximation is also derived via this argument, we briefly review it below; see \cite{wosencraft} for more details.


\subsection{Sphere packing}

\begin{figure}[t]
    \centerline{\includegraphics[width=3.3in]{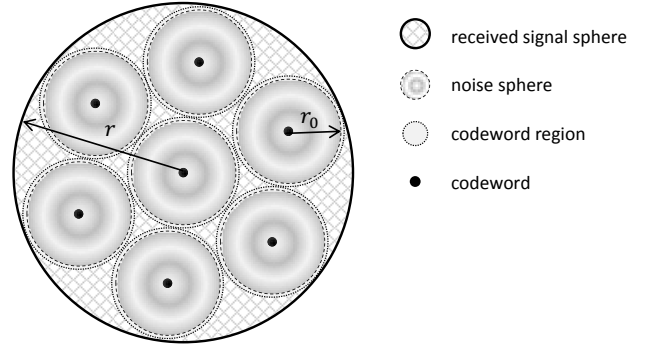}}
    \caption{Sphere-packing argument: a codeword region is just slightly larger than a corresponding noise sphere; $r_0\approx \sqrt{n\sigma_0^2}, r\approx \sqrt{n(\sigma_x^2+\sigma_0^2)}$.}
    \label{fig:2}
\end{figure}

During the transmission, $n$ consecutive symbols are grouped into codewords, so that $\bx = [x_1, x_2, ..., x_n]$ is transmitted while $\by = \bx + \bxi$ is received, where $\bxi = [\xi_1, \xi_2, ..., \xi_n]$ is the noise vector. As $n$ is getting large, the total noise power $|\bxi|^2=\sum_i \xi_i^2$ approaches $n\sigma_0^2$ with high probability (due to the law of large numbers), which is known as "sphere hardening" \cite{wosencraft}, so that $\by$ belongs to the noise sphere centered on a transmitted  codeword, as shown in Fig. \ref{fig:2}, with high probability. The decoder decides in favor of a particular codeword when received signal $\by$ belongs to the codeword region of that codeword (also known as "decoding region"). For any transmitted codeword, the received signal $\by$ belongs with high probability to the received signal sphere of radius $\sqrt{n(\sigma_x^2+\sigma_0^2)}$ which encloses all individual noise spheres centered on respective codewords. For reliable decoding, (i) codeword regions must enclose corresponding noise spheres, and (ii) they must not overlap. These can be satisfied by selecting codeword regions as spheres of a radius slightly larger than that of noise spheres and by packing all those spheres into the received signal sphere as tightly as possible allowing no overlaps.

The number $N$ of distinct codewords that can be reliably transmitted over the channel is approximately equal to the number of non-overlapping noise spheres that can fit into the received signal sphere. It can be estimated from the ratio of respective volumes:
\bal
N \approx V/V_0=\left(1+\gamma\right)^{n/2},
\eal
where $V= \alpha(n\sigma_x^2+n\sigma_0^2)^{n/2}$ and $V_0= \alpha(n\sigma_0^2)^{n/2}$ are the volumes of the received signal and noise spheres respectively, $\alpha = \pi^{n/2}/\Gamma(n/2+1)$ and $\Gamma(\cdot)$ is the gamma function, so that the channel capacity is
\begin{equation}
\label{eq.Ca}
C \approx \frac{1}{n}\log N \approx \frac{1}{2}\log(1+\gamma).
\end{equation}
Remarkably, this heuristic argument gives the exact value of the channel capacity for the AWGN channel.

\subsection{Achievable information rates}

The argument above applies if the number of codewords of lenght $n$ can be as large as necessary (i.e. is not constrained). For a fixed constellation of $M$ points, the number of codewords of length $n$ can be at most $M^n$. To evaluate the impact of this constraint, we present the per-symbol AIR in the following form:
\begin{equation}
\label{eq.CM.delta}
C_M = C - \Delta C,
\end{equation}
where $C$ is the  (unconstrained) channel capacity (as above) and $\Delta C \ge 0$ is the rate loss due to a fixed constellation of cardinality $M$. To estimate the latter, consider a hypothetical system with noise power $\sigma_1^2$ such that the number of distinct codewords (noise spheres) is exactly $M^n$:
\bal
M^n = \frac{V}{V_1}= \frac{(n\sigma_x^2+n\sigma_1^2)^{n/2}}{(n\sigma_1^2)^{n/2}},
\eal
where $V_1= \alpha(n\sigma_1^2)^{n/2}$ is the volume of the hypothetical noise sphere, which is also the volume of a codeword region when there are exactly $M^n$ codewords. For this system, there is no loss in capacity due to a fixed constellation (within the sphere packing approximation) since the noise power is "right" (i.e. noise spheres are the same as respective codeword regions so that no more codewords can fit without increasing the error probability):
\bal
\sigma_1^2 = \frac{\sigma_x^2}{M^2-1} \approx \frac{\sigma_x^2}{M^2},
\eal
where the last approximation holds when $M$ is reasonably large, $M^2 \gg 1$. However, if the true noise power is less than the hypothetical one, $\sigma_0^2 < \sigma_1^2$, more than one noise sphere can fit within the hypothetical noise sphere (codeword region), as shown in Fig. 3 (the central sphere). The resulting $\Delta C$ can be interpreted as the capacity of the fictitious  channel with signal power $\sigma_1^2$ and the noise power $\sigma_0^2$, which can be estimated via the ratio of volumes approach similarly to \eqref{eq.Ca}:
\bal
\Delta C \approx \frac{1}{n} \log \frac{(n\sigma_1^2+n\sigma_0^2)^{n/2}}{(n\sigma_0^2)^{n/2}} \approx \frac{1}{2}\log\left(1+\frac{\sigma_x^2}{M^2 \sigma_0^2}\right).
\eal
Substituting this in \eqref{eq.CM.delta}, one finally obtains an approximation $C_a$ for the per-symbol AIR of $M$-PAM:
\bal
\label{eq.CM.approx}
C_M \approx C_a = \frac{1}{2}\log \frac{1+\gamma}{1+\gamma/M^2}.
\eal
While this approximation is simple and analytically tractable, it is also reasonably accurate over the whole SNR range and for various constellation cardinalities $M$ (especially when $M$ is large), as Fig. \ref{fig:4} shows, being somewhat less accurate in the transition region between low and high SNR regimes and for 2-PAM (for which more accurate approximation is obtained in the next section). In all considered cases, the approximation inaccuracy does not exceed about 0.3 b/symb, which transforms into 30\% for 2-PAM, 10\% and 5\% for 8-PAM and 64-PAM, respectively. Hence, the approximation is more accurate for higher-rate constellations and also significantly more accurate outside of the transition region (between low and high SNR). Since the experimental results are well below the ideal AIR, this approximation is more accurate than the ideal AIR, with respect to the rates of real-world systems/experiments (see also the next section).

\begin{figure}[t]
    \centerline{\includegraphics[width=3in]{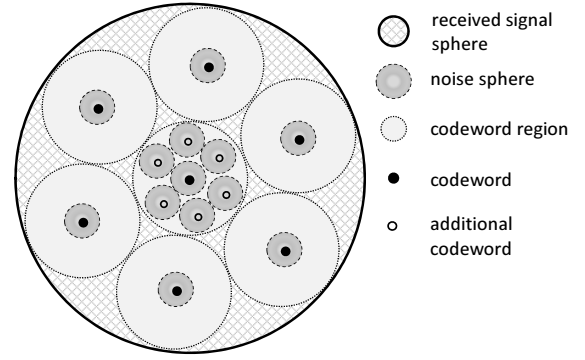}}
    \caption{The impact of the limited number of codewords: more noise spheres representing additional codewords could be packed into  existing codeword regions when noise is small.}
    \label{fig:3}
\end{figure}

\begin{figure}[t]
	\centerline{\includegraphics[width=3.7in]{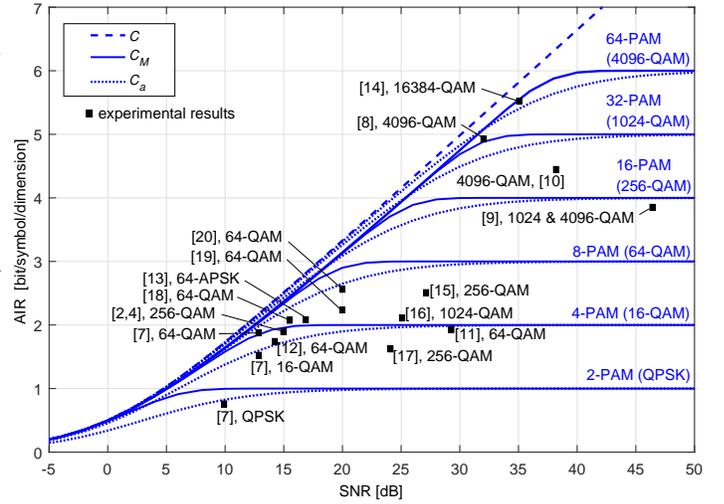}}
	\caption{M-PAM (QAM) AIR in [b/sym./dimension] evaluated via numerical integration in \eqref{eq.CM}, its approximation in \eqref{eq.CM.approx} (or \eqref{eq:MQAM}) and the AWGN channel capacity in \eqref{eq.C} vs. SNR; recent experimental results are also shown for comparison. }
	\label{fig:4}
\end{figure}

Not only this approximation is much simpler than its exact counterpart in \eqref{eq.CM} and thus facilitates a numerical or even analytical optimization (e.g. to find an optimal power allocation in a multi-stream transmission system, for which no closed-form solution is known \cite{lozano}), but it also captures important properties of the ideal AIR $C_M$ and provides additional insights unavailable from the known results:

1)  As $M$ increases, both $C_a$ and $C_M$ approach $C$: $C_a \approx C_M \approx C \ \mbox{if} \ M \gg \sqrt{\gamma}$, from which one can estimate minimum $M$ required to approach closely the channel capacity without using unnecessarily large constellations\footnote{This approximation demonstrates that the upper bound in \cite{ozarow}, which can be put in the form $M_{min} \le 2\sqrt{1+\gamma}$, is actually tight. Note also that the upper bound was obtained in \cite{ozarow} via an elaborate information-theoretic analysis (which does not yield a capacity approximation) while our approximation to $M_{min}$ follows directly from the approximation in \eqref{eq.CM.approx}.}:
\begin{equation}
\label{eq.Mmin}
		M_{min} \approx 2\max\{1,\sqrt{\gamma}\}.
\end{equation}
In this regime, the upper bound in \eqref{eq.CM.bound} is tight.

2) At high SNR, both $C_a$ and the ideal AIR $C_M$  approach the constellation entropy and are independent of the SNR,
	\begin{equation}
    \label{eq:CMa.high SNR}
		C_a \approx C_M \approx \log M \ \mbox{if} \ \gamma \gg M^2,
	\end{equation}
so that the upper bound in \eqref{eq.CM.bound} is also tight at high SNR.

3) At low SNR, $C_a$ can be approximated as follow:
	\begin{equation}
		C_a \approx\left(1-\frac{1}{M^2}\right)\frac{\log e}{2} \gamma \ \mbox{if} \ \gamma \ll 1,
	\end{equation}
so that the rate loss is small if $M^2 \gg 1$ and never exceeds 25\%. Comparing this to \eqref {eq:CMa.high SNR}, one concludes that the rate loss (due to constrained modulation) is additive at low SNR and multiplicative at high SNR, i.e. much more pronounced in the latter case.
	
4) The approximation $C_a$ is a continuously-differentiable (to any order) and increasing function of the SNR:
	\begin{equation}
		\frac{\partial C_a}{\partial \gamma}=\frac{\log e}{2}\frac{M^2-1}{(1+\gamma)(M^2+\gamma)}>0.
	\end{equation}

5) $C_a$ is a strictly-concave function of the SNR:
	\begin{equation}
		\frac{\partial^2 C_a}{\partial \gamma^2}=\frac{\log e}{2}\frac{(1-M^2)(1+M^2+ 2\gamma)}{(1+\gamma)^2(M^2+\gamma)^2}<0.
	    \label{eq:CMa.concave}
	\end{equation}

The latter two properties are instrumental in numerical optimization of  multi-stream transmission systems (e.g. MIMO, as in \cite{lozano}) by rendering the respective optimization problems convex so that all powerful tools of convex optimization, see e.g. \cite{boyd}, can be used.

\subsection{$M$-QAM}

While the approximation above was derived for the $M$-PAM constellation (which is 1-D), it can also be extended to $M$-QAM (2-D), which is a popular choice for modern optical transmission systems \cite{Karlsson-17}\cite{Cho-19} as well as wireless systems (e.g. WiFi, LTE, 5G). Indeed, $M$-QAM can be considered as 2 $\times$ $\sqrt{M}$-PAM operating on in-phase and quadrature channels \cite{barry} (here we assume that $\sqrt{M}$ is integer), so that its per-symbol AIR can be approximated as follows:
\begin{equation}
 	C_{M-QAM} = 2C_{\sqrt{M}-PAM} \approx \log\frac{1+\gamma}{1+\gamma/M}
	\label{eq:MQAM},
\end{equation}
and the properties above hold with the substitution $M \rightarrow \sqrt{M}$. In particular, the minimum constellation cardinality to approach closely the channel capacity is
\bal
\label{eq.Mmin.QAM}
M_{min} \approx 4 \max\{1,\gamma\}
\eal
Since 1st equality in \eqref{eq:MQAM} holds for both the approximations and ideal AIR, the relative accuracy of the $M$-QAM approximation is the same as that of $M$-PAM, see Fig. 4, where the rates are in [bit/symbol/dimension] (recall that $M$-PAM is 1-D while $M$-QAM is 2-D; experimental results are 4-D since two polarizations are used). Comparing this approximation with recent state-of-the art experimental results\footnote{It should be noted that the quoted experimental results were obtained under different conditions, e.g. single-channel vs. multi-channel systems etc., and hence differ significantly.} for back-to-back (B2B) or long-haul end-to-end (E2E) rates in Fig. \ref{fig:4}, one concludes that the approximation is in fact closer to the experimental rates than the ideal AIR. Note a particularly good agreement for QPSK, 16-QAM and 64-QAM in \cite{Cho-18}, and 64-QAM in \cite{Mazur-20}. We attribute this to the fact that the approximation $C_{M-QAM}$ in \eqref{eq:MQAM} underestimates the ideal AIR $C_M$ in \eqref{eq.CM} while the experimental B2B or E2E rates are also below the ideal AIR by about 0.5-2 [b/sym.] (due to finite blocklength/complexity codes with non-zero overhead and non-zero guard bands, in addition to other imperfections of practical systems \cite{Winzer-18}) Hence, the proposed approximations are more accurate than the ideal AIR $C_M$ with respect to the rates of real-world systems.

\section{An Asymptotic Approximation}
\label{sec:Integral}

In order to improve the accuracy of the AIR approximations above for BPSK and QPSK, we obtain here an approximation via the tools of asymptotic analysis (see e.g. \cite{Olver} for more details on these tools).  Applying the Laplace method to the integrals in \eqref{eq.CM}, the AIR of 2-PAM (BPSK) and 4-QAM (QPSK) can be approximated, after some manipulations, as
\begin{equation}
\label{eq.C2a}
	C_{2-PAM}\approx 1-\log\left(1+e^{-\gamma} \right),\ C_{4-QAM} = 2 C_{2-PAM}
\end{equation}
As Fig. 5 shows, this approximation is remarkably accurate over the whole SNR range: the inaccuracy does not exceed 10\%, being much smaller outside of the transition region. Note also that this approximation is a continuously-differentiable (to any order) and concave function of the SNR, making it suitable for convex optimization algorithms.

An alternative approximation for the 2-PAM (BPSK) and 4-QAM (QPSK) rates, which is somewhat more involved due to its use of the error function, can be found in \cite[(45), (46)]{Secondini-13}.

\begin{figure}[t]
	\centerline{\includegraphics[width=3in]{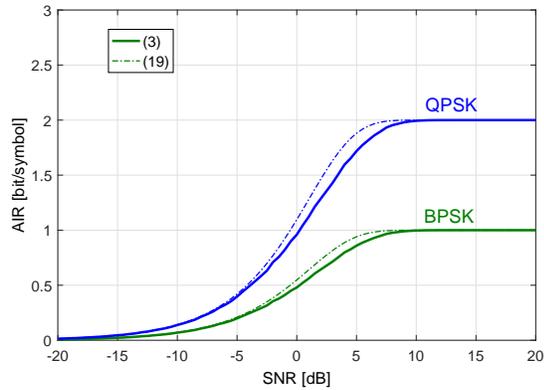}}
	\caption{2-PAM (BPSK) and 4-QAM (QPSK) AIR and the approximations in \eqref{eq.C2a} vs. SNR.}
	\label{fig:5}
\end{figure}

\section{Conclusion}
\label{sec:concl}
Based on the intuitive sphere-packing approach, the simple and  analytically-tractable approximations have been obtained for the achievable information rates of $M$-PAM and $M$-QAM constellations over the AWGN channel, including optical fiber channels when nonlinearity can be modeled by additive Gaussian noise. Not only these approximations are reasonably accurate over the whole SNR range and for various values of $M$ (including large ones), they also provide additional insights unavailable from the known expressions. The sphere-packing based derivation sheds additional light on the rate loss mechanism due to a fixed (constrained) constellation cardinality: while the loss is additive at low SNR, it is multiplicative, i.e. much more pronounced, at high SNR.

Comparison to the B2B or E2E rates of state-of-the art real systems/experiments shows that the proposed simple approximations are in fact more accurate than the ideal AIR.

The presented approach and approximations can also be used for  polarization-multiplexed transmission systems. Since the difference between the MI and GMI is small for Gray-labeled constellations \cite{Alvarado-18},  the above approximations can also be used for the latter.


\begin{IEEEbiographynophoto} {Maria Urlea} received her B.A.Sc. (2012) and M.A.Sc (2015) in Telecommunications form the School of Electrical Engineering and Computer Science at the University of Ottawa. She is currently a Systems Engineer in Ottawa, Canada and focuses on network design and architectures for several local and national entities.

\end{IEEEbiographynophoto}

\begin{IEEEbiographynophoto} {Sergey Loyka} 	was born in Minsk, Belarus. He received the Ph.D. degree in Radio Engineering from the Belorussian State University of Informatics and Radioelectronics (BSUIR), Minsk, Belarus in 1995 and the M.S. degree with honors from Minsk Radioengineering Institute, Minsk, Belarus in 1992. Since 2001 he has been a faculty member at the School of Electrical Engineering and Computer Science, University of Ottawa, Canada. Prior to that, he was a research fellow in the Laboratory of Communications and Integrated Microelectronics (LACIME) of Ecole de Technologie Superieure, Montreal, Canada; a senior scientist at the Electromagnetic Compatibility Laboratory of BSUIR, Belarus; an invited scientist at the Laboratory of Electromagnetism and Acoustic (LEMA), Swiss Federal Institute of Technology (EPFL), Lausanne, Switzerland. His research areas are wireless communications and networks and, in particular, MIMO systems and security aspects of such systems, in which he has published extensively. He received a number of awards from the URSI, the IEEE, the Swiss, Belarus and former USSR governments, and the Soros Foundation.
\end{IEEEbiographynophoto}

\end{document}